\pdfoutput=1

\documentclass[conference]{IEEEtran}

\def\shownotes{1}  

\usepackage{courier}
\usepackage{amsmath}
\usepackage{amssymb}

\usepackage[]{graphicx}
\usepackage{subfigure}
\usepackage{epsfig}
\usepackage{epstopdf}

\usepackage{algorithm,algorithmic}
\usepackage{boxedminipage}
\usepackage{multirow}
\usepackage{tabularx}
\usepackage{comment}
\usepackage{color}

\usepackage{enumitem}
\setlist{nolistsep}

\providecommand{\ie}{\emph{i.e.,} }
\providecommand{\eg}{\emph{e.g.,} }

\providecommand{\myparab}[1]{\smallskip\noindent\textbf{#1} }

\newtheorem{lemma}{Lemma}

\newtheorem{proposition}{Proposition}

\ifnum\shownotes=1
\newcommand{\authnote}[2]{{ $\ll$\textsf{\footnotesize #1 notes: #2}$\gg$}}
\else
\newcommand{\authnote}[2]{}
\fi

\begin{document}

\title{
Mind Your Own Bandwidth:
An Edge Solution to Peak-hour Broadband Congestion
}


\author{
\IEEEauthorblockN{Felix Ming Fai Wong\IEEEauthorrefmark{1}, Carlee Joe-Wong\IEEEauthorrefmark{1},
Sangtae Ha\IEEEauthorrefmark{2}, Zhenming Liu\IEEEauthorrefmark{1}, Mung Chiang\IEEEauthorrefmark{1}}
\IEEEauthorblockA{\IEEEauthorblockA{\IEEEauthorrefmark{1}Princeton University, Princeton, NJ, USA
\IEEEauthorrefmark{2}University of Colorado, Boulder, CO, USA}
\{mwthree, cjoe, chiangm\}@princeton.edu, sangtae.ha@colorado.edu,zhenming@cs.princeton.edu}

}

\maketitle

\begin{abstract}
Motivated by recent increases in network traffic,
we propose a decentralized network edge-based solution to peak-hour broadband congestion that
incentivizes users to moderate their bandwidth demands to their actual needs.
Our solution is centered on smart home gateways that allocate bandwidth in a two-level hierarchy:
first, a gateway purchases guaranteed bandwidth from the Internet Service Provider (ISP) with virtual credits. It then
self-limits its bandwidth usage and distributes the bandwidth among its apps and devices according to their relative priorities.
To this end, we design a credit allocation and redistribution mechanism for the first level,
and implement our gateways on commodity wireless routers for the second level.
We demonstrate our system's effectiveness and practicality
with theoretical analysis, simulations and experiments on real traffic. Compared to a baseline equal sharing algorithm, our solution significantly improves users' overall satisfaction and yields a fair allocation of bandwidth across users.

\end{abstract}

\IEEEpeerreviewmaketitle


\section{Introduction}

\subsection{Motivation: Demand in Cable Networks}
In recent years, ISPs have seen a large, sustained increase in traffic demand on their wired and wireless networks, driven by the increasing popularity of media streaming services such as Netflix and cloud services such as Dropbox \cite{CiscoVNI}. While many strategies for managing this demand have been proposed recently \cite{Sen}, few works explicitly consider wired cable networks.
Cable providers have themselves considered two measures to manage network congestion:
1) protocol/content-agnostic fair bandwidth distribution, where fairness can account for a household's recent usage history \cite{congestioncomcast},
and 2) using deep packet inspection to detect and throttle ``abusive" users, \eg those running BitTorrent.
However, these existing solutions cannot simultaneously address the following two challenges:

\textbf{Incentivizing responsible usage.}
Fair sharing does not directly translate to user satisfaction, \eg when everyone overloads the network by streaming HD videos at the same time.
The crux of the current problem is that \emph{users have no incentive to moderate their bandwidth consumption only to the amount they actually need}.
We need a solution that accounts for peak-hour usage over longer timespans, \eg one week, so that users can plan their usage patterns accordingly.

\textbf{Addressing different bandwidth needs.}
When bandwidth is a limited resource it is important to prioritize certain types of traffic while maintaining
privacy and net neutrality.
Moreover, users should have a way to express their usage preferences to the ISP.
Consider, for instance, two neighbors who both use a substantial amount of bandwidth in the evening. One of them watches Netflix, and the other backs up large files from work. The optimal solution would then be to prioritize the first neighbor during the evening, and provide an incentive to the second neighbor to back up his files a few hours later at night.
Yet with current network infrastructure,
the cable provider might sub-optimally allocate equal bandwidth to both neighbors when congestion is present.

\subsection{A Home Solution to a Home Problem}\label{sec:solution}

We argue the difficulties faced by cable providers can be solved by pushing congestion management to the network edge
at home gateways. Thus, we \emph{empower the user} to improve his or her own satisfaction from using the network.

We propose to allocate bandwidth using a two-level hierarchy, as shown in Figure \ref{fig:hierarchy}. On the first level, bandwidth is allocated among home gateways. At a second level, each gateway's bandwidth is allocated among the users and devices connected to the gateway.

Central to our solution on Level 1 is the notion of \emph{virtual credits}, inspired by~\cite{Kelly-PFP}, distributed to the gateways. Each gateway uses its credits to ``purchase'' guaranteed bandwidth rates at congested times, and thus has an incentive to moderate network usage due to its limited credit budget. We limit the total bandwidth demand to the network capacity by fixing the total number of credits available to spend and recirculating credits to the gateways as they are spent. Using credits enables us to meet the following requirements:

\textbf{Fairness:} Credits are circulated back to each gateway in a way that depends on other gateways' behavior. Over time, every gateway will be able to use a fair portion of the bandwidth, as gateways that spend a lot of credits in one time period will have fewer to spend later.

\textbf{Social welfare optimization:} At the equilibrium, each gateway chooses its credit spending so as to selfishly maximize its own satisfaction. We show that the credit redistribution mechanism ensures that these choices also optimize the collective social welfare, \ie \emph{all} gateways' satisfaction, over time. 

\textbf{Decentralization:} The circulation of credits in our system naturally allows for a distributed solution to the equilibrium, since each gateway decides how to spend its credits. We design an algorithm for making these spending decisions that runs at individual gateways and utilizes \emph{only the information on how the credits are circulated} to find the optimal spending.

\textbf{Privacy preservation:} Since gateways' spending decisions use only their knowledge of the credit recirculation, they need not reveal their individual preferences to other gateways or to the ISP, thus preserving their privacy. 

\textbf{Incremental deployability:} Since the number of total credits is fixed, we can incrementally deploy the solution by starting with a small number of credits, and introducing more as more gateways begin to participate. 

Once each gateway has chosen a bandwidth rate, we perform a second-level allocation, dividing this gateway's total bandwidth among the devices and applications. Different apps and different users can then receive more or less bandwidth depending on their relative priorities (\eg video streaming over software updates).

\subsection{System Architecture}

\begin{figure}
\centering
\includegraphics[width=0.45\textwidth]{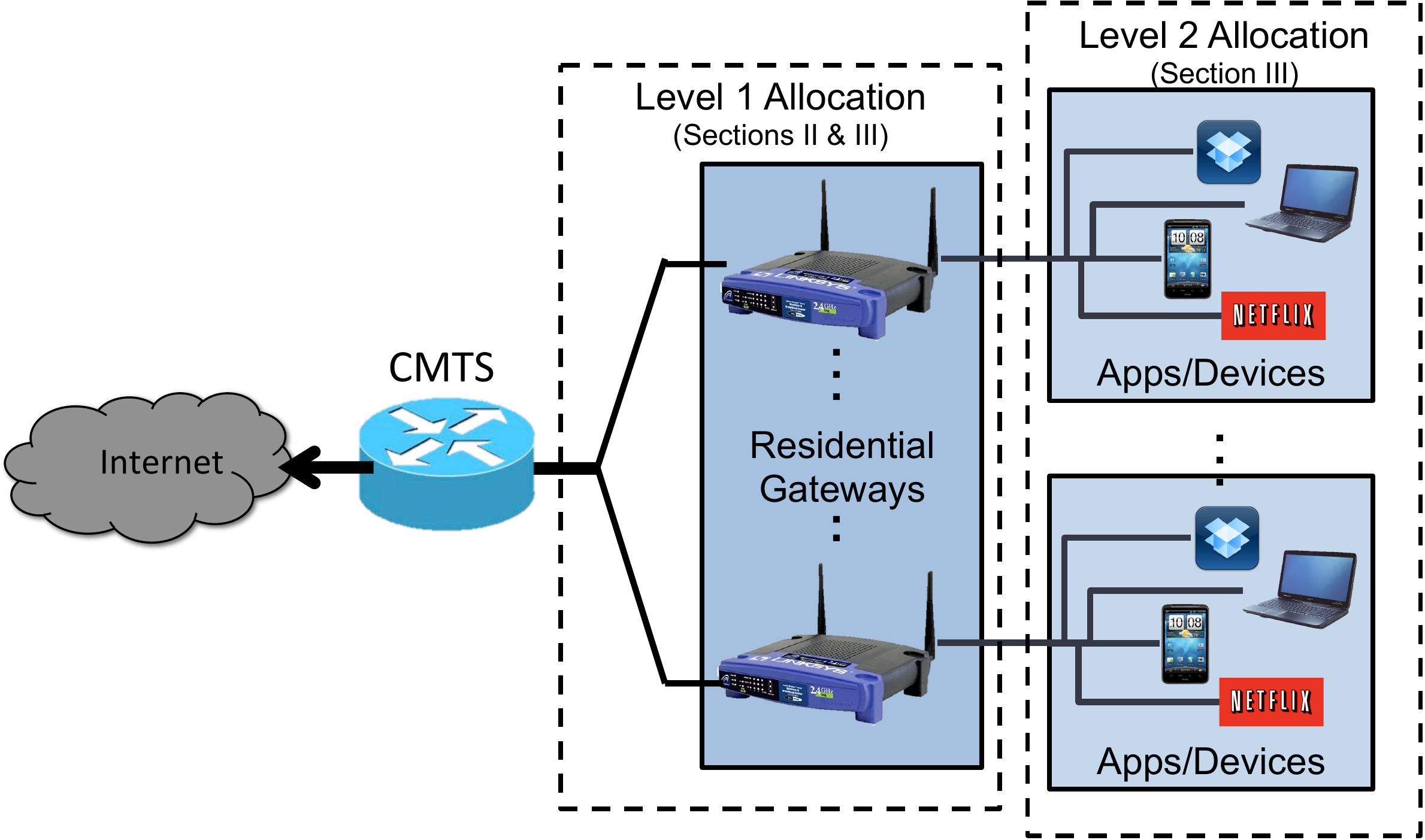}
\vspace{-0.05in}
\caption{Hierarchical bandwidth allocation.}
\vspace{-0.15in}
\label{fig:hierarchy}
\end{figure}

The architecture of our proposed solution is shown in Figure~\ref{fig:hierarchy}. We consider a series of discrete time periods (\eg each lasting one hour) and allocate bandwidth among gateways and users in each period. At the start of each period, the algorithm performs a two-level allocation: first, each gateway decides how many of its credits to spend, \ie how much guaranteed bandwidth to purchase in this period (Level 1). In practice, an automated agent acting on behalf of the gateway's users will make this decision, though some users may wish to manually override the gateway's decision. Once this decision is made, the gateway performs the second-level allocation, dividing the purchased bandwidth among its apps and devices.

In each time period, a central server in the ISP's network records the total credits spent by each gateway and redistributes the appropriate number of credits to each gateway in the next time period. Each gateway updates its budget by deducting the credits spent in this time period and adding the number of credits redistributed to it. In the next time period, each gateway then knows its updated budget and can again choose how many credits to spend.

\myparab{Level 1.} When congestion is present, the ISP divides traffic into two classes: a first-tier, higher priority class, which gateways must purchase with credits, and a second-tier class that costs no credits but is always of lower priority.
The first traffic class provides users with a guaranteed minimum bandwidth, which can be utilized by other gateways' second-tier traffic only when there is spare capacity. This scheme ensures that the network is fully utilized if there is sufficient demand, yet still eases congestion through encouraging gateways to spend credits at different times. Moreover, our scheme allows an infinite variety of service classes for first-tier traffic, each defined by its guaranteed bandwidth rate. Second-tier traffic is the lowest tier, as it has no guaranteed rate. By using credits, our algorithm accomplishes this multi-class allocation without introducing substantial overhead for the cable provider.

We discuss the Level 1 credit allocation in Sections \ref{sec:algorithm} and \ref{sec:utilities} of the paper. Section \ref{sec:algorithm} introduces the credit redistribution algorithm and characterizes the optimal solution, \ie the spending pattern that maximizes the gateways' total satisfaction. Section \ref{sec:utilities} then presents a practical online algorithm for each gateway to independently decide the amount of credits to spend at each time, subject to predictions about the number of credits redistributed in the future.

\myparab{Level 2.} Our level 2 allocation mechanism distributes each gateway's bandwidth share among its applications according to their different priorities. In Section \ref{sec:utilities}, we give an algorithm for optimally distributing bandwidth, and we discuss its implementation in Section \ref{sec:implementation}.
In performing this allocation, we emphasize two properties that ensure its practicality:

\textbf{User-specified device/app prioritization:} Each user has different priorities for their devices--one user, for instance, might prioritize streaming music, while another might prioritize file transfers. Given these priorities, we automate the bandwidth allocation among the different devices and apps.\footnote{Should explicit prioritization prove too complex for average users, we can introduce default priorities for different types of apps and devices.} We focus on \emph{elephant traffic}, which tends to be non-bursty and amenable to bandwidth throttling.

\textbf{Network neutrality:} The level 2 allocation runs locally at each gateway. Thus, the user has full control of these decisions, maintaining ISP neutrality.

\noindent We make the following contributions in this paper:
\begin{itemize}
\item A virtual pricing mechanism to fairly allocate limited network capacity among gateways and incentivize users to limit their bandwidth demand, easing congestion.
\item A distributed algorithm that allows users to optimally choose their bandwidth used.
\item A gateway implementation that accordingly limits the total bandwidth used. 
\item A practical method for classifying traffic at the gateway and enforcing bandwidth limits on different apps. 
\end{itemize}

After discussing our bandwidth allocation algorithms in Sections \ref{sec:algorithm} and \ref{sec:utilities}, we describe our implementation in Section \ref{sec:implementation} and show simulation and implementation results of an example scenario in Sections \ref{sec:experiments} and \ref{sec:simulations}. We briefly discuss related works in Section \ref{sec:related} and conclude in Section \ref{sec:conclusion}.

\section{Credit Distribution and Optimal Spending} \label{sec:algorithm}

In this section, we describe the bandwidth allocation at the higher level of Figure~\ref{fig:hierarchy}. We first describe our system of credits for purchasing bandwidth (Section \ref{sec:isp}) and show that it satisfies several fairness properties. We then show that even if each gateway selfishly maximizes its own satisfaction, the total satisfaction across all gateways can be maximized (Section \ref{sec:global}). All proofs are in Appendices \ref{sec:proof1}--\ref{sec:prooflast}.

\subsection{Credit Distribution}\label{sec:isp}
We divide congested times of the day into several discrete time periods, e.g., of a half-hour duration, and allow gateways to ``purchase'' bandwidth in each time period. At the end of the period, spent credits are redistributed to the gateways.

We suppose that a fixed number $B = \beta C$ of credits is shared by $n$ different gateways, where $C$ is the network capacity in Gbps and $\beta$ an over-provisioning factor chosen by the ISP. We consider $T + 1$ time periods indexed by $t = 0,1,\ldots,T$, e.g., $T + 1$ periods per day. We use $b_{it}$ to denote the budget of each gateway $i$ at time $t$, and we suppose that the total credits are initially distributed equally across gateways, i.e., $b_{i0} = B/n$ for all $i$. We then update each gateway $i$'s budget as
\begin{equation}
b_{i,t + 1} = b_{it} - x_{it} + \sum_{j \neq i}\frac{x_{jt}}{n - 1},
\label{eq:credits}
\end{equation}
where we sum over all gateways except gateway $i$ and $x_{it}$ denotes the number of credits used by gateway $i$ in time period $t$. Each gateway $i$ is constrained by $0\leq x_{it} \leq b_{it}$: it cannot spend negative credits, and the number of credits spent cannot exceed its budget. This credit redistribution scheme conserves the total number of credits for all times $t$:
\begin{lemma}\label{lem:conserve}
At any time $t$, the number of credits  distributed among gateways is fixed, i.e., $\sum_{i = 1}^n b_{it} = B = \beta C$.
\end{lemma}

Heavy gateways are prevented from hogging the network (helping to enforce a form of \emph{fairness across gateways}), as a large $x_{jt}$ simply means that the other gateways $i\neq j$ will receive larger budgets in the time interval $t + 1$. In fact, if this redistribution leads back to a previous budget allocation, then all gateways spend the same number of credits:
\begin{lemma}\label{lem:equal}
Suppose that for some times $s$ and $t$, $b_{is} = b_{it}$ for all gateways $i$, e.g., $s = 0$ and $b_{it} = B/n$. Then each gateway spends the same number of credits between times $s$ and $t$: for all gateways $i$ and $j$,
$\sum_{\tau = s}^{t - 1} x_{i\tau} = \sum_{\tau = s}^{t - 1} x_{j\tau}$.
\end{lemma}

Using this result, we can more generally bound the difference in the number of credits gateways can spend:
\begin{proposition}\label{prop:bound}
At any time $t$, for any two gateways $i$ and $j$, $\left|\sum_{s = 0}^t x_{is} - \sum_{s = 0}^t x_{js}\right| \leq B(n - 1)/n$. Thus, the time-averaged difference in spending
\begin{equation}
\lim_{t\rightarrow\infty}\frac{1}{t}\left|\sum_{s = 0}^t x_{is} - \sum_{s = 0}^t x_{js}\right| \leq \lim_{t\rightarrow\infty}\frac{B(n - 1)}{nt} = 0.
\end{equation}
\end{proposition}
Over time, fairness is enforced in the sense that all gateways can spend approximately the same number of credits.

Though these fairness results to some extent limit heavy gateways' hogging the network, lighter gateways may conversely ``hoard'' credits, thus hurting other gateways' budgets. Yet no single gateway can hoard all available credits:
\begin{proposition}\label{prop:limit}
Suppose that a given gateway $i$ uses at least $\epsilon$ bandwidth every $p$ periods, where $\frac{B}{n} > \epsilon \geq 0$ and $p$ may denote, e.g., one day. Then at any time $t$, gateway $i$'s budget
\begin{align}
b_{it}&\leq \frac{B}{n}\alpha^{t + 1} + B\left(1 - \alpha^{t + 1}\right) - \epsilon\left(\frac{\alpha^p - \alpha^{p\left(1 + \left\lfloor\frac{t + 1}{p}\right\rfloor\right)}}{1 - \alpha^p}\right) \nonumber \\
&\rightarrow B - \epsilon\left(\frac{(n - 2)^p}{(n - 1)^p - (n - 2)^p}\right) \label{eq:limit}
\end{align}
as $t\rightarrow\infty$, where $\alpha = (n - 2)/(n - 1)$. In particular, if $\epsilon > 0$, $b_{it} < B$. Moreover, at any fixed time $t$, at most one gateway can have a budget of zero credits.
\end{proposition}
For instance, if a gateway spends $\epsilon$ credits at each time, then as $t\rightarrow\infty$, $b_{it} \leq B - \epsilon(n - 2)$: if $\epsilon$ is relatively large, a gateway hoards fewer credits, since these are redistributed among others once spent. Conversely, a gateway that spends very little can asymptotically hoard almost $B$ credits. 

More broadly, if a number $m$ of gateways are inactive in a network for a certain number of time periods $s$, then we can bound the number of credits these $m$ gateways accumulate:
\begin{proposition}\label{prop:limit_num}
Suppose that $m$ gateways are inactive for $s$ time periods. Then the number of credits that these gateways can accumulate in these times is given by
\begin{equation}
\sum_{i = 1}^m b_{is} - b_{i0} \leq \left(1 - \left(\frac{n - 2}{n - 1}\right)^s\right)\left(B - \sum_{i = 1}^m b_{i0}\right)
\end{equation}
where we index the inactive gateways by $i = 1,2,\ldots,m$ and suppose they are inactive from time 0 to time $s - 1$.
\end{proposition}
Thus, gateways can only accumulate all the credits asymptotically, as $s\rightarrow\infty$. In practice, however, a gateway might stay inactive for a long period of time, hoarding credits and decreasing other gateways' utilities. Thus, we cap each gateway's budget at a maximal value of $\overline{B}$ to limit hoarding.

To ensure that gateways can still save some credits for future use, we choose $\frac{B}{n} \leq \overline{B}\leq B$. Since $\sum_{i = 1}^n b_{it} = B$ at each time $t$, this lower bound ensures that there is a feasible set of budgets $\left\{b_{it}\right\}$ with each $b_{it} \leq \overline{B}$.\footnote{If $\overline{B} = B/n$, then we would have $b_{it} = B/n$ for all gateways $i$ at all times $t$, so in practice we would have $(B/n) < \overline{B} < B$.} For instance, the ISP might choose $\overline{B} = \frac{B}{n - m + 1}$, where $m$ equals the minimum number of gateways on the network at any given time. The $n - m$ inactive gateways at that time can then hoard at most $B(n - m)/(n - m + 1)$ credits, leaving the remaining $B/(n - m + 1)$ credits for the active gateways.

To enforce this budget cap, the excess budget
$\left(b_{it} - x_{it} + \sum_{j \neq i}\frac{x_{jt}}{n - 1}\right) - \overline{B}$
of any gateway $i$ exceeding the cap is evenly distributed among all gateways below the cap. Should these excess budgets push any gateway over the cap, the resulting excess is evenly redistributed to the remaining gateways until all budgets are below the cap. Since we choose $\overline{B} > \frac{B}{n}$ and reallocate to fewer gateways at each successive iteration, this process converges after at most $n - 1$ iterations.

\subsection{Optimal Credit Spending}\label{sec:global}

Given the above credit distribution scheme, each gateway must decide how many credits to spend in each period. To formalize this mathematically, let $U_{it}$ denote gateway $i$'s utility as a function of the guaranteed bandwidth $x_{it}$ in time interval $t$.  Though gateways may increase their utilities with second-tier traffic, we do not consider this traffic in our formulation. Second-tier bandwidth is difficult to predict, as most gateways would not have consistent historical information on its availability. Gateways could only learn this information by sending such traffic, which they might not do regularly.

We consider a finite time horizon $T$ to keep the problem tractable and because the utility functions cannot be reliably known indefinitely far into the future. Each gateway $i$ then optimizes its total utility from the current time $s$ to $s + T$:
\begin{equation}
\max_{x_{it}} \sum_{t = s}^{s + T} U_{it}\left(x_{it}\right),\;\;{\rm s.t.}\;0\leq x_{it} \leq b_{it},\;\forall t.
\label{eq:gateway_opt}
\end{equation}
Here the budgets $b_{it}$ are calculated using the credit redistribution scheme (\ref{eq:credits}), with appropriate adjustments to enforce a budget limit. For ease of analysis, we do not model these budget limits here. In practice, the ISP can cap gateways' budgets for each time period during the credit redistribution.

We first note that the budget expressions (\ref{eq:credits}) can be used to rewrite the inequality $x_{it} \leq b_{it}$ as the linear function
\begin{equation}
\sum_{\tau = s}^t x_{i\tau} - \sum_{j\neq i}\sum_{\tau = s}^{t - 1} \frac{x_{j\tau}}{n - 1} \leq b_{i0},
\label{eq:budget_new}
\end{equation}
Thus, if the $U_{it}$ are strictly concave functions, then given the amount spent by other gateways $x_{j\tau}$, (\ref{eq:gateway_opt}) is a convex optimization problem with linear constraints.\footnote{The assumption of concavity, i.e., $U_{it}''(x_{it}) < 0$, may be justified with the economic principle of diminishing marginal utility as bandwidth increases.}

Since each gateway chooses its own $x_{it}$ to solve (\ref{eq:gateway_opt}), these joint optimization problems may be viewed in a game-theoretic sense: each gateway is making a decision that affects the utilities of other gateways. From this perspective, the game has a Nash equilibrium at the system optimum:
\begin{proposition}\label{prop:nash}
Consider the global optimization problem
\begin{equation}
\max_{x_{it}} \sum_{i = 1}^n \sum_{t = s}^{s + T} U_{it}\left(x_{it}\right),\;\;{\rm s.t.}\;0\leq x_{it} \leq b_{it},\;\forall i,t
\label{eq:isp_opt}
\end{equation}
with the credit redistribution (\ref{eq:credits}) and strictly concave $U_{it}$. Then an optimal solution $\left\{x_{it}^\ast\right\}$ to (\ref{eq:isp_opt}) is a Nash equilibrium.
\end{proposition}

While Prop. \ref{prop:nash}'s result is encouraging from a system standpoint, in practice this Nash equilibrium may never be achieved. Since the gateways do not know each others' utility functions, they do not know how many credits will be spent and redistributed at future times, making the future credit budgets unknown parameters in each gateway's optimization problem. These must be estimated based on historical observations, which we discuss in the next section.

\section{An Online Bandwidth Allocation Algorithm}\label{sec:utilities}

In this section, we consider a gateway's actions at both levels of bandwidth allocation. We first give an algorithm to decide its credit spending (Level 1), and then show how the purchased bandwidth can be divided among apps at the gateway (Level 2). Using Algorithm \ref{alg:allocate}, each gateway iteratively estimates the future credits redistributed, decides how many credits to spend, prioritizes apps, and updates its credit estimates. We assume throughout that the gateway's automated agent knows the utility functions for users at that gateway. 

\setlength{\textfloatsep}{0pt}
\begin{algorithm}[t]
\small
\caption{Gateway spending decisions.}
\label{alg:allocate}
\begin{algorithmic}
\STATE $s\leftarrow 1$ \COMMENT{$s$ tracks the current time.}
\WHILE{$s > 0$}
	\IF{$s > 1$}
		\STATE Update estimate of future amounts redistributed using Algorithm \ref{alg:prediction}.
	\ENDIF
	\STATE Calculate $\sum_{j \neq i}x_{jt}/(n - 1)$ for $t = s,\ldots, s + T - 1$.
	\STATE Solve (\ref{eq:gateway_opt}) with budget constraints (\ref{eq:budget_new_exp}) given $\sum_{j \neq i}x_{jt}/(n - 1)$.
	\STATE Choose the application priorities $\mu_k$ by solving (\ref{eq:fairdevice}).
	\STATE $s\leftarrow s + 1$
\ENDWHILE
\end{algorithmic}
\end{algorithm} 

\subsection{Estimating Other Gateways' Spending}\label{sec:est_other}

To be consistent with (\ref{eq:gateway_opt})'s finite time horizon, we suppose that gateways employ a \emph{sliding window} optimization. At any given time $s$, gateway $i$ chooses rates for the next $T$ periods $s,\ldots, s + T -1$ so as to maximize its utility for those periods. At time $s + 1$, the gateway updates its estimates of future credits redistributed and optimizes over the next $T$ periods, etc.

Estimating the future credits spent by other gateways is difficult: since relatively few gateways share each cable link, fluctuations in a single gateway's behavior can significantly affect the number of credits redistributed.
Thus, we propose the method of \emph{scenario optimization} to estimate the number of credits each gateway will receive in the future. This technique is often used in finance to solve optimization problems with stochastic constraints that are not easily predicted, e.g., due to market dynamics \cite{consiglio2007scenario}.

Scenario optimization considers a finite set $S_i$ of possible scenarios for each gateway $i$ and computes the optimal spending $x_{it}$ in each scenario. Each scenario $\sigma\in S_i$ is associated with a probability $\pi_\sigma$ that the scenario $\sigma$ will take place. Evaluating the credit redistribution for each $\sigma$ then yields a probability distribution of the possible credits spent.

In our case, a ``scenario'' can be defined as a set of utility functions $\left\{U_{jt}\right\}$ for other gateways. We can parameterize these scenarios by noting that gateways' utilities depend on the application used, e.g., streaming versus downloading files. We incorporate this dependence by taking
\begin{equation}
U_{jt} = \gamma_{jt}\left(p_{jt}^1 u_1 + p_{jt}^2 u_2 + p_{jt}^3 u_3 + p_{jt}^4 u_4\right),
\label{eq:utility}
\end{equation}
for each gateway $j$, where $\gamma_{jt}$ is a scaling factor and each $u_k$ is the utility received from an application of type $k$ (e.g., $k = 1$ corresponds to streaming, $k = 2$ to file downloads, etc.) The utilities $u_k$ are assumed to be pre-determined functions consistent across gateways, and the $\gamma_{it}$ are specified by individual gateways. The variable $p_{jt}^k$ corresponds to the (estimated) probability that gateway $j$ optimizes its usage using the utility function $u_k$, e.g., if application $k$ is used the most at time $t$. This probabilistic approach accounts for a gateway's incomplete knowledge of its future utility functions.

With this utility definition, we can define a scenario $\sigma$ by the coefficients $\gamma_{jt}(\sigma)$ and $p_{jt}^{k}(\sigma)$ of gateways' utility functions. Since gateway $i$ has no way to distinguish between other gateways, it can group them together as one ``gateway'' $j$ by simply adding the utility functions and budget constraints. These aggregated gateways $j$ then maximize
\begin{equation*}
U_{jt} = \sum_{t = 1}^T\gamma_{jt}(\sigma)\sum_{k = 1}^4 \left(p_{jt}^k(\sigma) u_k\right)
\end{equation*}
subject to the budget constraints $0\leq x_{jt} \leq b_{jt}$, where the coefficients $\gamma_{jt}(\sigma)p_{jt}^k(\sigma)$ represent the added coefficients for all gateways $\neq i$.
We suppose that gateway $j$ correctly estimates gateway $i$'s future usage $x_{it}$, and that gateway $i$ also correctly estimates $x_{jt}$. Thus, following Prop. \ref{prop:nash}, all gateways choose their usage so as to maximize the collective utility $\sum_t \left(U_{jt} + U_{it}\right)$ subject to the budget constraints. This optimization may be solved to calculate the credits $\sum_{j\neq i} x_{jt}/(n - 1)$ redistributed at each time $t$ in scenario $\sigma$.

To improve our credit estimates, at each time $t$ we update the scenario probabilities $\pi_\sigma$ by comparing the observed number of credits redistributed at time $t - 1$, denoted by $\sum_{j\neq i} \overline{x}_{j,t - 1}/(n - 1)$, with the estimated amount redistributed $\sum_{j\neq i} x_{j,t - 1}(\sigma)/(n - 1)$ for each $\sigma\in S_i$. We suppose that gateways' behavior is sufficiently periodic such that the scenario probabilities at times $t$ and $t + T$ are the same.

We use $P\left(\sum_{j\neq i} \overline{x}_{j,t - 1} = \sum_{j\neq i} x_{j,t - 1}(\sigma)\right)$ to denote the probability that, given $\sum_{j\neq i} \overline{x}_{j,t - 1}/(n - 1)$ at time $t$, gateways use scenario $\sigma$'s utility function at time $t$. We can calculate these probabilities by measuring the $L_2$ discrepancy between the estimated and observed credits redistributed:
\begin{align*}
P&\left(\sum_{j\neq i} \overline{x}_{j,t - 1} = \sum_{j\neq i} x_{j,t - 1}(\sigma)\right) = \\
&\frac{1}{\left|S_i\right| - 1}\left(1 - \frac{\left(\sum_{j\neq i} \overline{x}_{j,t - 1} - \sum_{j\neq i} x_{j,t - 1}(\sigma)\right)^2}{\sum_{l = 1}^{|S_i|} \left(\sum_{j\neq i} \overline{x}_{j,t - 1} - \sum_{j\neq i} x_{j,t - 1}(l)\right)^2}\right).
\end{align*}
We then update the scenario probabilities $p_\sigma$ using Bayes' rule and use
the new $p_\sigma$ in Algorithm \ref{alg:prediction}.
\setlength{\textfloatsep}{0pt}
\begin{algorithm}[t]
\small
\caption{Estimating credit redistribution.}
\label{alg:prediction}
\begin{algorithmic}
\STATE $s\leftarrow 1$ \COMMENT{$s$ tracks the current time.}
\WHILE{$s > 0$}
	\FORALL[this loop may be run in parallel]{gateways $i = 1,\ldots,n$}
		\STATE Choose scenarios $S_i$.
		\FOR{each scenario $\sigma\in S$}
			\STATE Calculate the predicted amount redistributed $\sum_{j \neq i}x_{jt}(\sigma)/(n - 1)$ for $t = s,\ldots, s + T - 1$, assuming other gateways know $x_{it}$ for all $t$.
			\IF{$s > 1$}
				\STATE Update probability $p_\sigma$ using Bayes' Rule.
			\ENDIF
		\ENDFOR
	\ENDFOR
\ENDWHILE
\end{algorithmic}
\end{algorithm}

\subsection{Online Spending Decisions and App Prioritization}\label{sec:online}

Algorithm \ref{alg:allocate} shows how the credits spent in different scenarios are incorporated into choosing a gateway's rates $x_{it}$ and application priorities. Each gateway constrains its spending depending on the estimated redistributed credits: for instance, a conservative gateway might choose the $x_{it}$ so that the budget constraints $0\leq x_{it} \leq b_{it}$ hold for all scenarios. In the discussion below, we suppose that gateways ensure that the constraints hold for the expected number of credits redistributed; the budget constraint (\ref{eq:budget_new}) then becomes
\begin{equation}
\sum_{\tau = s}^t x_{i\tau} - \sum_{\sigma\in S_i}p_\sigma\left(\sum_{j\neq i}\sum_{\tau = s}^{t - 1} \frac{x_{j\tau}(\sigma)}{n - 1}\right) \leq b_{i0}.
\label{eq:budget_new_exp}
\end{equation}
We additionally introduce the constraints $b_{it} \leq \overline{B}$, i.e., that the expected budget at a given time cannot exceed the budget cap. This constraint ensures that gateways are not forced to give credits to other gateways due to the cap.

Each gateway can further improve its own experience through its Level 2 allocation dividing the purchased bandwidth among its apps. It does so by assigning priorities to different devices and applications, so that higher-priority apps receive more bandwidth. Since users cannot be expected to manually specify priorities in each time period, we introduce an automated algorithm that leverages the gateway's known utility functions (\ref{eq:utility}) to optimally set application priorities.

We consider the four application categories in (\ref{eq:utility}) and use $\mu_k$ to represent each category $k$'s priority. Since the particular applications active at a given time may change within a given period, e.g., if a user starts or stops watching a video, we define an application's priority in relative terms: for any apps $k_1$ and $k_2$, $\mu_{k_1}/\mu_{k_2} = y_{k_1}/y_{k_2}$, where $y_k$ is the amount of bandwidth allocated to application $k$ and $\sum_k y_k = x_{it}$, ensuring that all the purchased bandwidth is used. We normalize the priorities to sum to 1, i.e., $\sum_k \mu_k = 1$.

Since it is nearly impossible to predict which apps will be active at a given instant of time, we choose the app priorities $\mu_k$ according to a ``worst-case scenario,'' in which all apps are simultaneously active. In this case, each app $k$ receives $\mu_k x_{it}$ bandwidth, and we choose the $\mu_k$ to maximize total utility:
\begin{equation}
\max_{\mu_k} \sum_{j = 1}^4 u_k\left(\mu_k x_{it}\right),\;\;{\rm s.t.}\;\sum_{k = 1}^m \mu_k = 1.
\label{eq:fairdevice}
\end{equation}
Since each function $u_k$ is assumed to be concave and the constraint is linear in the $\mu_k$ , (\ref{eq:fairdevice}) is a convex optimization problem and may be solved rapidly with standard methods.

\section{Design and Implementation}\label{sec:implementation}

\subsection{System Architecture}

\begin{figure}
\centering
\includegraphics[width=0.5\textwidth]{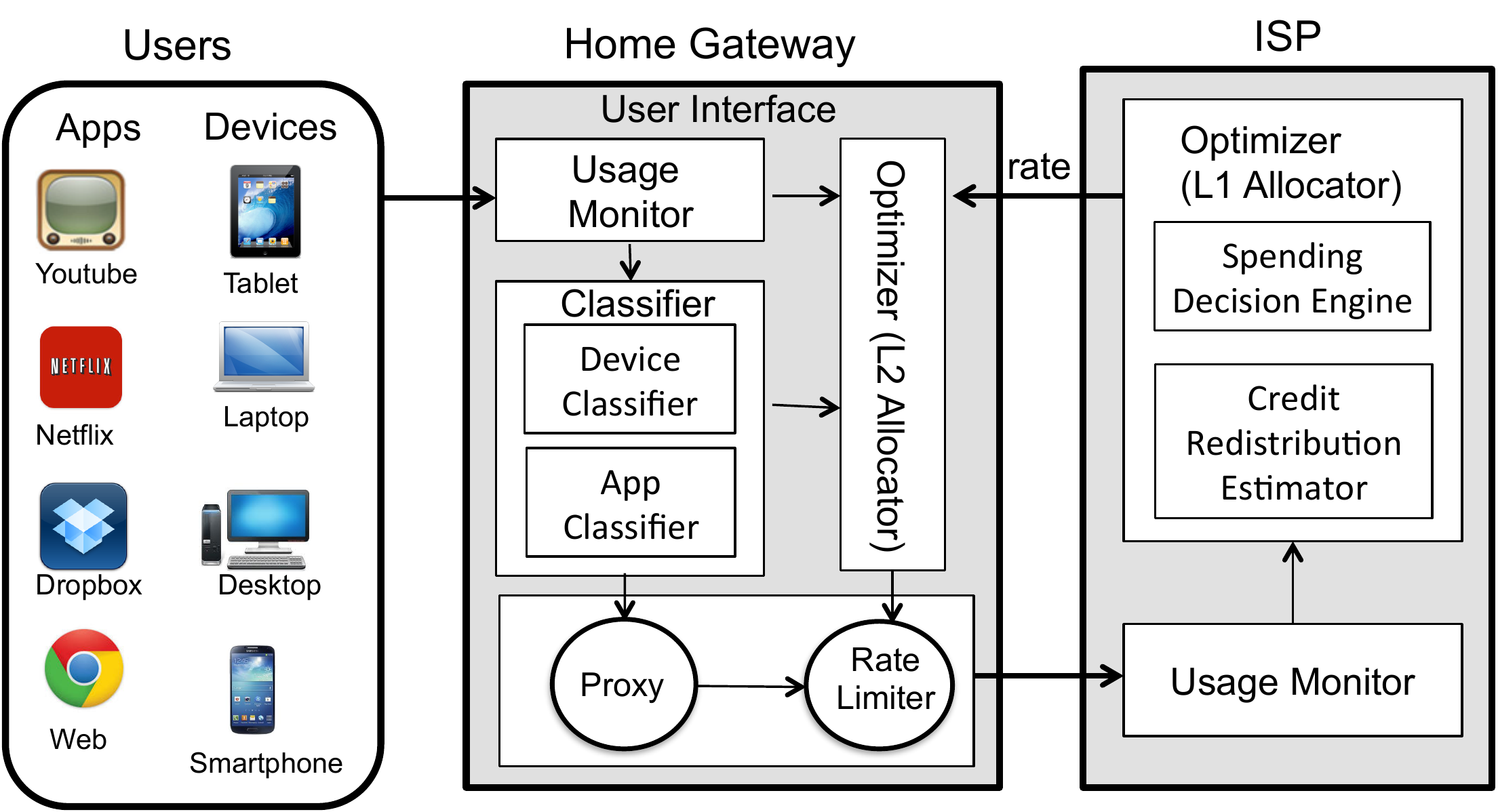}
\caption{System architecture.}
\label{fig:arch_block}
\end{figure}

The architecture of our system is summarized in Figure \ref{fig:arch_block}.
It consists of four modules:
1) When traffic goes through the gateway for forwarding, it is passed to 
a device and an application classifier to identify its traffic type and priority.
2) All traffic is redirected through a proxy process, which forwards traffic
between client devices and the Internet. The proxy's data forwarding rate is  
determined by the optimizer (L2 Allocator) in each gateway by considering an 
application's priority. The rate is enforced by a rate limiter.
3) The bandwidth (credit spending) for each gateway is computed by the optimizer 
(L1 Allocator) in the ISP.
4) A user can access the gateway through a web interface to view its usage
(at either aggregate or joint device-app level) and update its preferences, \ie
when to spend more credits and traffic priorities,
so as to adjust the optimizer's decisions. We show screenshots of these user interfaces in Figures \ref{fig:screenshot1} and \ref{fig:screenshot2}.
In this section we discuss how we implement the classification, rate limiting, 
and prioritization in a commodity router.

\subsection{Hardware and Software}
\begin{figure*}[htp]
\centering
\subfigure[Cisco E2100L board.]{
\includegraphics[width=3.5cm, height=4.6cm]{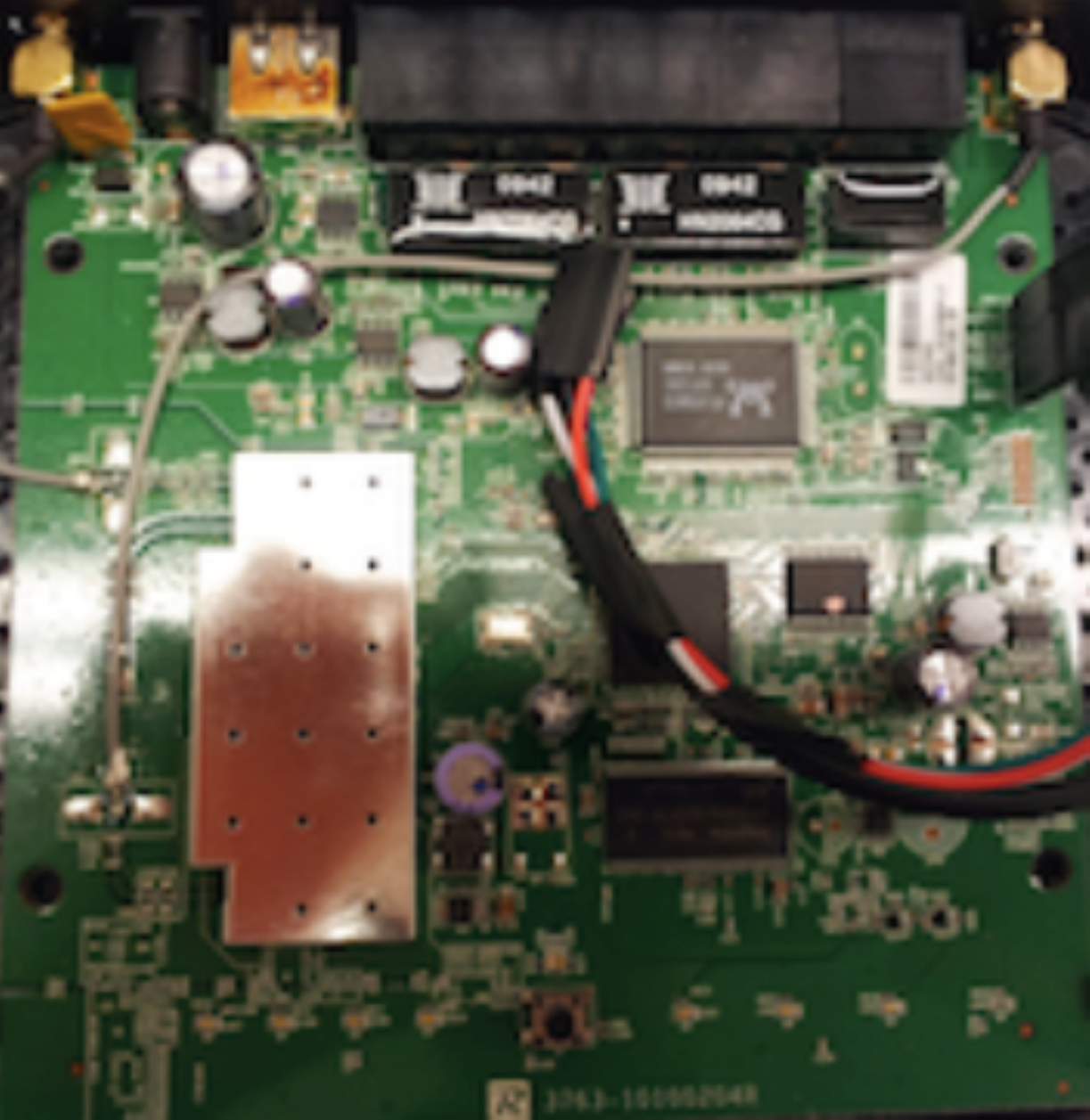}
\label{fig:hardware}
}
\subfigure[Usage tracking.]{
\includegraphics[width=6.6cm]{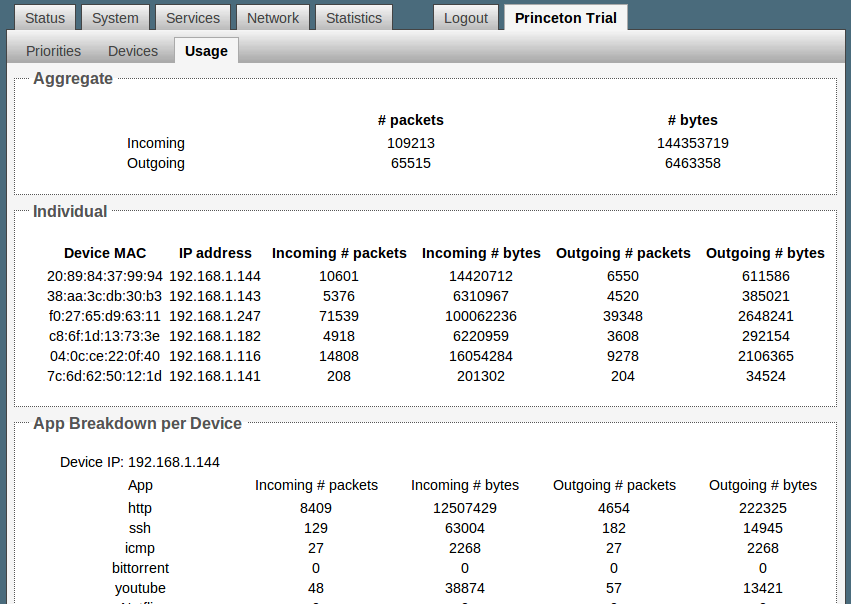}
\label{fig:screenshot1}
}
\subfigure[Traffic prioritization and device/OS classification.]{
\includegraphics[width=6cm]{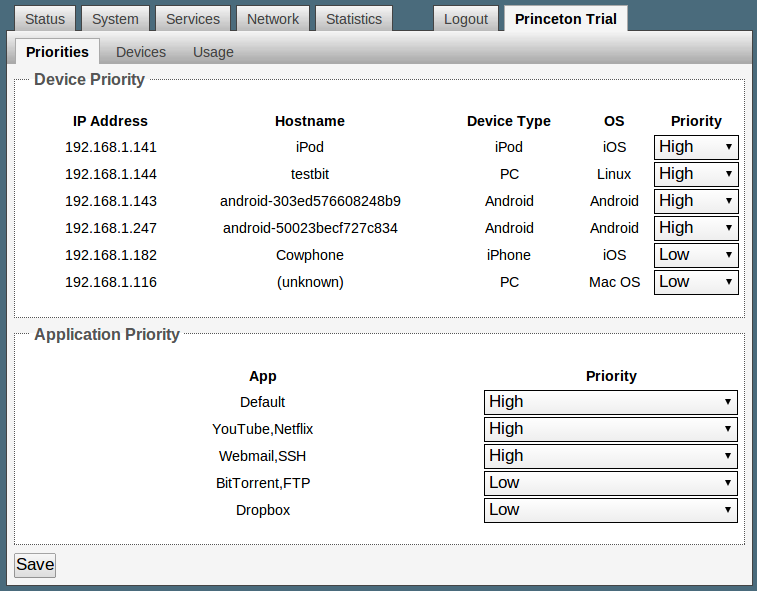}
\label{fig:screenshot2}
}
\caption{Router and web interface screenshots.}
\end{figure*}

We implement our system in a commodity wireless router, a Cisco E2100L with 
an Atheros 9130 MIPS-based 400MHz processor, 64MB memory, and 8MB flash storage (Figure \ref{fig:hardware}).
We replaced the factory default firmware with \mbox{OpenWrt}, a Linux
distribution commonly used for embedded devices.
Although OpenWrt has rich functionality in terms of traffic classification and
control, we still find it insufficient for our needs.
In particular, we modify OpenWrt to enable: 1) low-overhead traffic and device
classification using HTTP connections, \ie not port-based, and 2) back-pressure
based gentle rate limiting that does not mandate parameter tuning.

\subsection{Traffic and Device Classification}
To build a low-overhead classifier, we integrate a  kernel-level $\mathtt{netfilter}$ module
that inspects the first several packets of a connection for application matching.
If a match is found, the classifier marks the connection with a mark
to be queried at userspace by our proxy processes through $\mathtt{netlink}$.

Our classifier module performs traffic classification \emph{above} layer 7, \ie it can differentiate 
YouTube and Netflix, through a combination of content matching, byte tracking and protocol fingerprinting.
For device and OS classification, we use the same module to monitor HTTP traffic and inspect 
user-agent header strings for device information.
In practice we find this approach effective because of the prevalence of devices using HTTP traffic.

\subsection{Rate Limiting}
Our goal is to: 1) enforce an aggregate rate limit over
multiple connections, and 2) enforce prioritization, \ie which gets higher bandwidth,
among the connections given the aggregate limit.
Achieving this is challenging because we are throttling \emph{incoming}
traffic,\footnote{Throttling outgoing traffic is easy with token bucket-based
traffic shaping mechanisms.}
and at a packet level this is possible only by dropping packets
when the incoming rate exceeds a threshold (known as traffic policing),
or by delaying packet forwards to match
a specified rate limit through ACK clocking.
Since both approaches unnecessarily increase RTT and degrade user experience,
we instead take advantage of TCP's flow control mechanism and implement
our own rate limiting system in the application layer,
which is both easier to implement and more \emph{graceful} in throttling.
The module consists of two components.

\textbf{Transparent Proxy.}
When a connection between a client device and a server is being established, it is intercepted
at the gateway and redirected to the proxy process running in the gateway.
Then the proxy establishes a new connection to the server on behalf of the client
and forwards traffic between the two (proxy-server and client-proxy) connections.
We use the Linux $\mathtt{splice()}$ function to achieve zero-copying, \ie all data are handled in kernel space.

\textbf{Implicit Receive Window Control.}
TCP's flow control mechanism allows the receiver of a connection to advertise a
receive window to the sender so that incoming traffic does not overwhelm the receiver's buffer.
While originally set to match the available receiver buffer space,
the receive window can be artificially set to limit bandwidth using the
relation $\mathtt{cwnd} = \mathtt{rate} \times \mathtt{RTT}$: given a maximum 
$\mathtt{rate}$ and measured round trip time ($\mathtt{RTT}$), we set the receive window to be
no greater than $\mathtt{rate} \times \mathtt{RTT}$.
Although it is possible to set the receive window directly by modifying TCP headers, we opt
for a more elegant \emph{adaptive} approach such that the proxy does not need to know
the $\mathtt{RTT}$ or compute the exact window size.

To illustrate our approach we first consider a one connection case. As data from the
server arrives at the proxy, they are queued at the proxy's receive buffer until
the proxy issues a $\mathtt{recv()}$ on the proxy-server socket to process and clear them
(at the same time the proxy issues a $\mathtt{send()}$ on the client-proxy socket to
forward the data to the client). 
Note that if we modulate the freqeency and the size of $\mathtt{recv()}$'s, we
modulate the size of the receive buffer and effectively the sending rate.
Further explanation is given in Appendix \ref{sec:buffer_model}.

\subsection{Traffic Prioritization.}
When there are multiple connections the proxy spawns multiple threads such that each thread
serves one connection, and we aim to limit the aggregate rate $R$ over all connections.
To allocate bandwidth fairly among the connections, we coordinate socket reads of these threads
through a time division multiplexing scheme: using a thread mutex,
we create a virtual time resource such that each socket read is associated with an
exclusively held time slot of length proportional to the number of bytes read.
Although more complicated socket read scheduling mechanisms can be considered,
for simplicity we leave the scheduling to the operating system, and from experiments we
observe the sharing of time slots to be fair.

For traffic prioritization we assign a relative priority parameter $\alpha_i \in (0,1]$
for every connection $i$ such that for $n$ \emph{busy} connections, \ie each has
a sufficiently large backlog,
we want the sum of their rates $R_i$ to be $\sum_{i=1}^n R_i = R$, and
$R_i / R_j = \alpha_i / \alpha_j$ for $i,j=1,\ldots,n$.

We achieve the desired prioritization through truncated reads. When the proxy
issues a socket read, it needs to specify a maximum block size $b$ to read,\footnote{We
set it to be the page size of the processor architecture.}
and for a busy connection this limit $b$ is always reached.
If connection $i$ is of lower priority with $\alpha_i<1$, we truncate this block limit
by setting it to be $\alpha_i b$. Since each access to a time slot is associated with a
server socket read (equivalently, a client socket write) of $\alpha_i b$ bytes
and time slots are fairly distributed across connections,
the achieved client rate $D_i$ (equivalent to $R_i$) scales with $\alpha_i$.

By virtue of statistical multiplexing, our rate allocation mechanism does not require
the number of busy connections $n$, which is difficult to track in practice; hence it
can readily accommodate new connections.
To accommodate bursty connections, the proxy first queries the receive buffer for
the number of pending bytes. If it is above $b$ then it does a truncated read as
described above; otherwise it does not.
The pseudocode of a proxy thread is shown in Algorithm \ref{alg:ratelimit}.

\renewcommand{\algorithmicrequire}{\textbf{Input:}}
\begin{algorithm}[t]
\small
\caption{\small Pseudocode of incoming rate control.}
\label{alg:ratelimit}
\begin{algorithmic}
\REQUIRE $R$, $b$, $\alpha_i$,\\ $\mathtt{server\_fd}$: socket of server connection,\\ $\mathtt{client\_fd}$: socket of client connection,\\ $\mathtt{mutex}$: thread mutex shared by all connections
\WHILE{connection open}
\STATE $\mathtt{bytes\_read} = \mathtt{recv(server\_fd, }\ b \mathtt{)}$
\STATE $\mathtt{bytes\_per\_write} = \alpha_i \times \mathtt{bytes\_read}$
	\WHILE{not all $\mathtt{bytes\_read}$ written to client}
	\STATE $\mathtt{send(client\_fd, bytes\_per\_write)}$
	\STATE $\mathtt{lock(mutex)}$
	\STATE $\qquad \mathtt{sleep(bytes\_per\_write} / R)$
	\STATE $\mathtt{unlock(mutex)}$
	\ENDWHILE
\ENDWHILE
\end{algorithmic}
\end{algorithm}

\section{Experimental Results}\label{sec:experiments}

\begin{figure*}[htp]
\centering
\subfigure[Throughput.]{
\includegraphics[width=5cm]{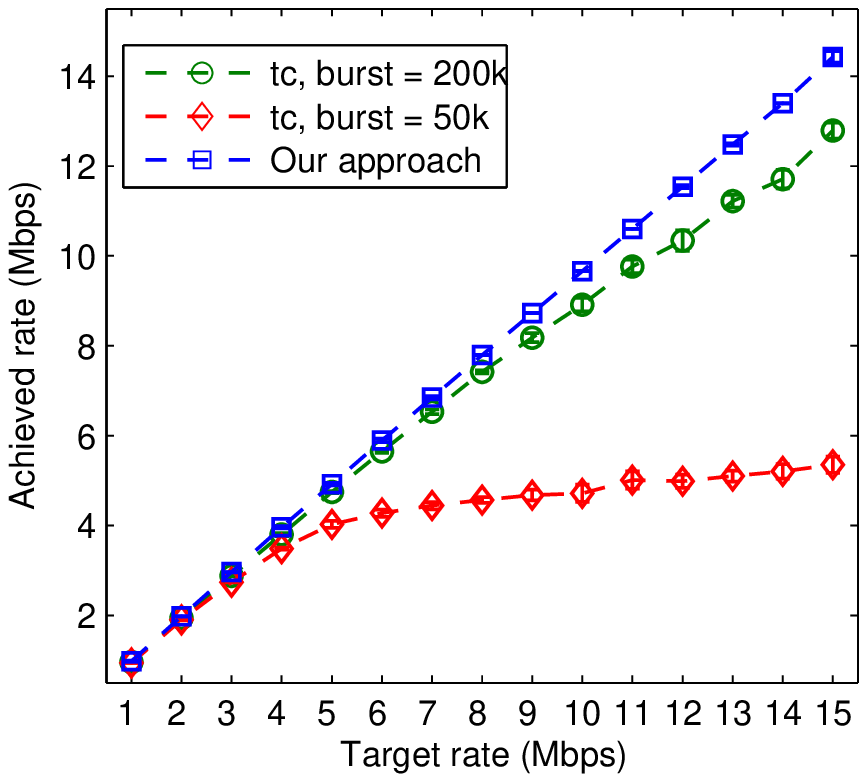}
\label{fig:compare_rate}
}
\subfigure[Retransmission counts.]{
\includegraphics[width=5cm]{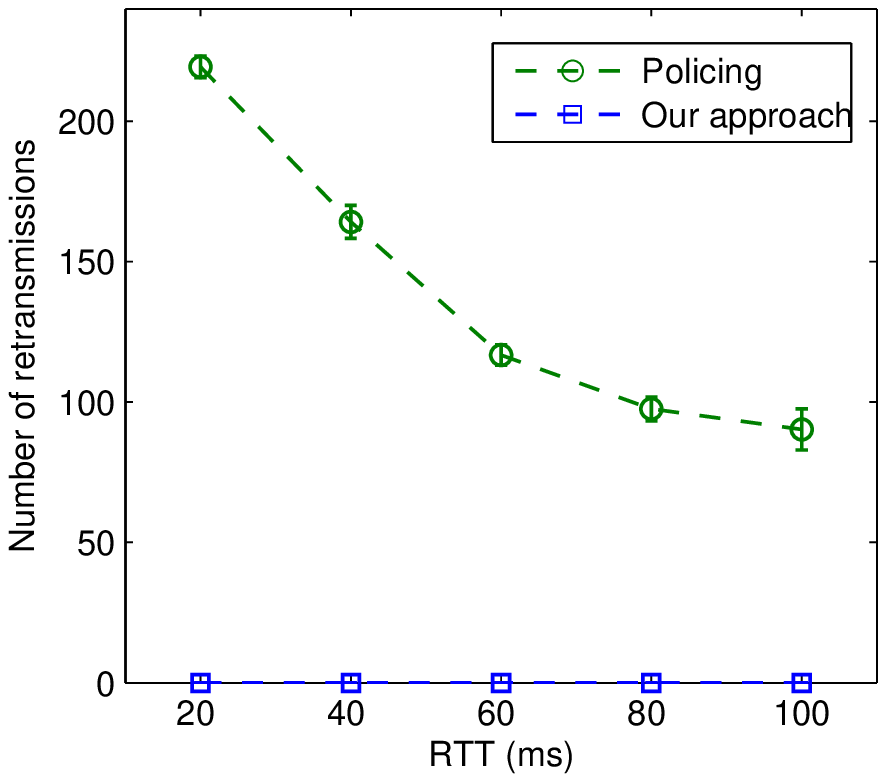}
\label{fig:compare_retrans}
}
\subfigure[Jitter.]{
\includegraphics[width=5cm]{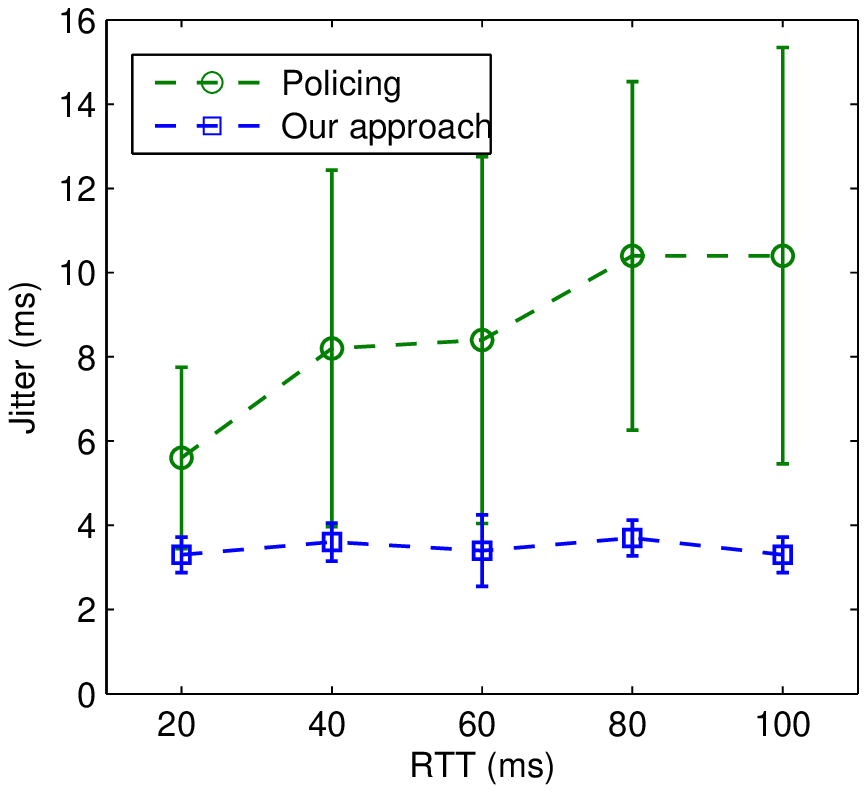}
\label{fig:compare_jitter}
}
\vspace{-0.1in}
\caption{Comparison between our rate limiting algorithm and $\mathtt{tc}$. Results shown are averaged over 10 runs, 60 seconds each, with 95\% confidence intervals.}
\vspace{-0.1in}
\end{figure*}

\subsection{Rate Limiting}
To justify our decision to develop our own rate control module, we compare it with the standard Linux traffic policing
approach using the $\mathtt{tc}$ command.
Two experiments are performed using $\mathtt{iperf}$.
In the first one, we fix network RTT to be 100ms and vary the rate limit from 1 to 15Mbps
to observe the actual rate achieved. We also try two choices of $\mathtt{tc}$'s burst parameter to ensure completeness.
Figure \ref{fig:compare_rate} shows that our approach results in more accurate rate limiting (less than 4\% error in each setting).
While it appears that increasing the burst parameter helps in improving rate limiting accuracy,
we note the values chosen are rather large (a typical value is 10k, while we use 50k and 200k) and may harm network stability.
The sensitivity of the results of $\mathtt{tc}$ w.r.t.\ parameters also highlights the need of careful parameter tuning,
which is undesirable given the diversity of network environments.

The first experiment hints that traffic policing, or using packet drops to signal the sender to reduce
its rate, is too drastic as a rate control mechanism. Our second experiment confirms this.
We fix the rate limit at 8Mbps and burst parameter at 50k, and vary network RTT from 20 to 100ms.
Figures \ref{fig:compare_retrans} and \ref{fig:compare_jitter} show that $\mathtt{tc}$ results in significantly more
packet retransmissions and higher jitter. This shows our approach is indeed more graceful in rate limiting.

\subsection{Traffic Prioritization}

\begin{figure}
\centering
\includegraphics[width=0.4\textwidth]{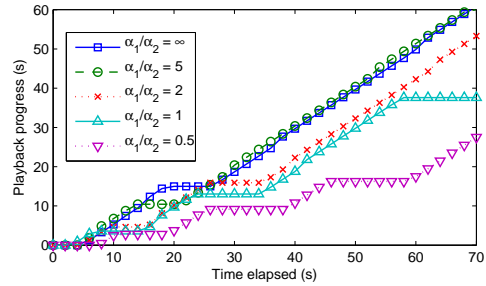}
\caption{YouTube playback performance with different prioritization settings.}
\label{fig:youtube_wget_prio}
\end{figure}

Consider a scenario with two users, one watching a 720p YouTube video stream
and the other downloading a large file with $\mathtt{wget}$, competing for a limited bandwidth of 2Mbps.
We vary the priorities of the two types of traffic and observe the effect on video playback.

Let $\alpha_1$ and $\alpha_2$ be the priorities of YouTube and $\mathtt{wget}$ respectively.
With $\alpha_1$ fixed, we vary $\alpha_2$ and measure the amount of video played w.r.t.\ time elapsed.\footnote{
We create a video-embedded webpage with a Javascript snippet that periodically queries the YouTube API for playback progress.}
Note there are two base cases: the case $\alpha_1/\alpha_2 = \infty$ corresponds to YouTube traffic
not interfered by $\mathtt{wget}$ and is the best possible result we can expect;
the case $\alpha_1/\alpha_2 = 1$ is equivalent to not having prioritization.
Figure \ref{fig:youtube_wget_prio} shows the results. When $\alpha_1/\alpha_2 < 1$, i.e., YouTube has
higher priority, playback performance (inversely related to the duration of pauses or flat regions in a curve)
is strictly better than the no prioritization case. Also, performance improves with decreasing $\alpha_1/\alpha_2$
ratio.

Not only is our system able to do fine-grained traffic classification with two types of traffic running under HTTP, but our traffic prioritization algorithm also produces noticeable improvement in user experience.

\section{Gateway Sharing Simulation}\label{sec:simulations}

To demonstrate the efficacy of our gateway sharing framework, we consider sixteen gateways sharing a cable link. After explaining the simulation setup, we compare our credit-based allocation to \emph{equal sharing}, in which the ISP reduces all gateways' bandwidth to a minimal but acceptable level, e.g., 1 Mbps. With equal sharing, each gateway is assigned a slot that receives this bandwidth until the network capacity is reached. This approach, which is similar to cable operators' current practices in that gateways are all treated equally, has two problems.
First, it risks inefficiency: gateways may occupy a slot without needing the slot's full bandwidth. Second, gateways that need more bandwidth cannot receive any from gateways that do
not. Our credit-based approach addresses both disadvantages, and we show in our simulations that it significantly improves gateway utilities. Moreover, all gateways receive a fair rate allocation, with no one gateway receiving significantly more bandwidth than the others.

After comparing the credit-based and equal sharing solutions, we consider the solution obtained with the online algorithm \ref{alg:allocate} in Section \ref{sec:utilities}. Despite the gateways' not fully knowing their future credit budgets, the utility achieved is near to the optimal. Moreover, all gateways receive a fair share of bandwidth, and gateways actively save and spend credits at different times.

\subsection{Gateway Utilities and Simulation Parameters}

\begin{figure*}
\centering
\vspace{-0.1in}
\subfigure[Jain's Index for gateway rates.]{
\includegraphics[width = 0.39\textwidth]{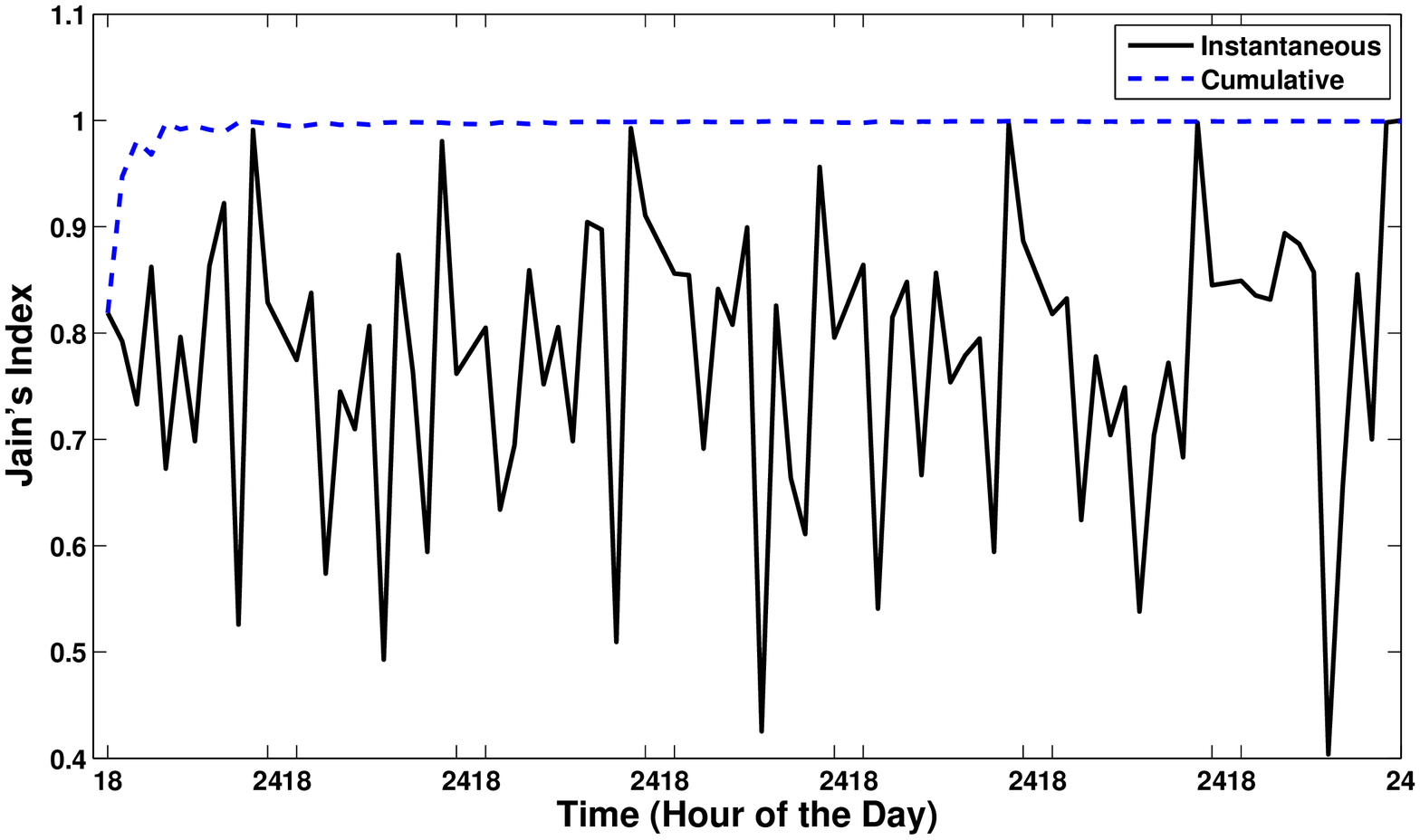}
\label{fig:rates_opt}}
\hspace{-0.05\textwidth}
\subfigure[Credit budgets for representative gateways.]{
\includegraphics[width = 0.39\textwidth]{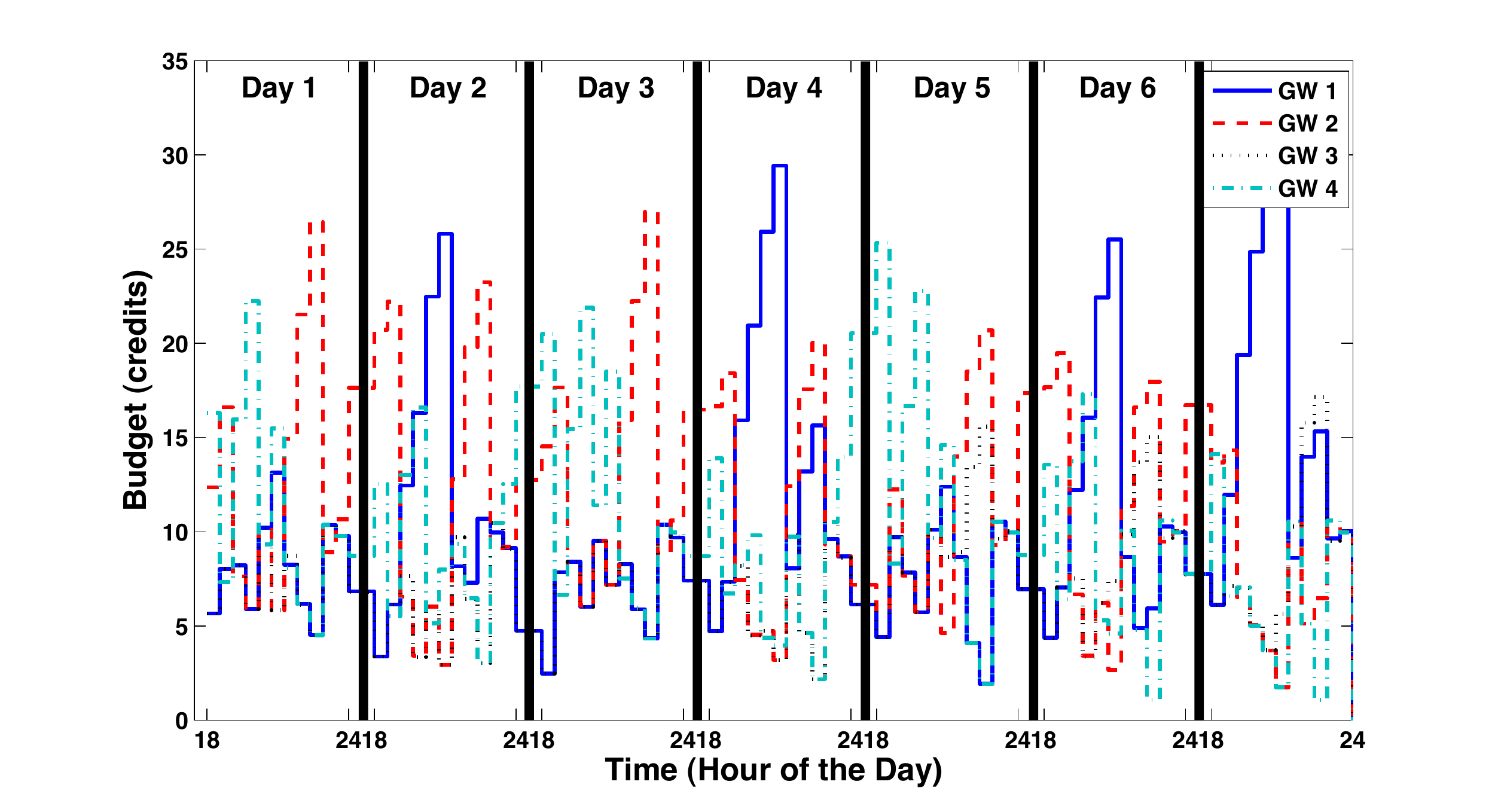} 
\label{fig:budgets_opt}}
\hspace{-0.045\textwidth}
\subfigure[Utility comparison.]{\includegraphics[width = 0.26\textwidth]{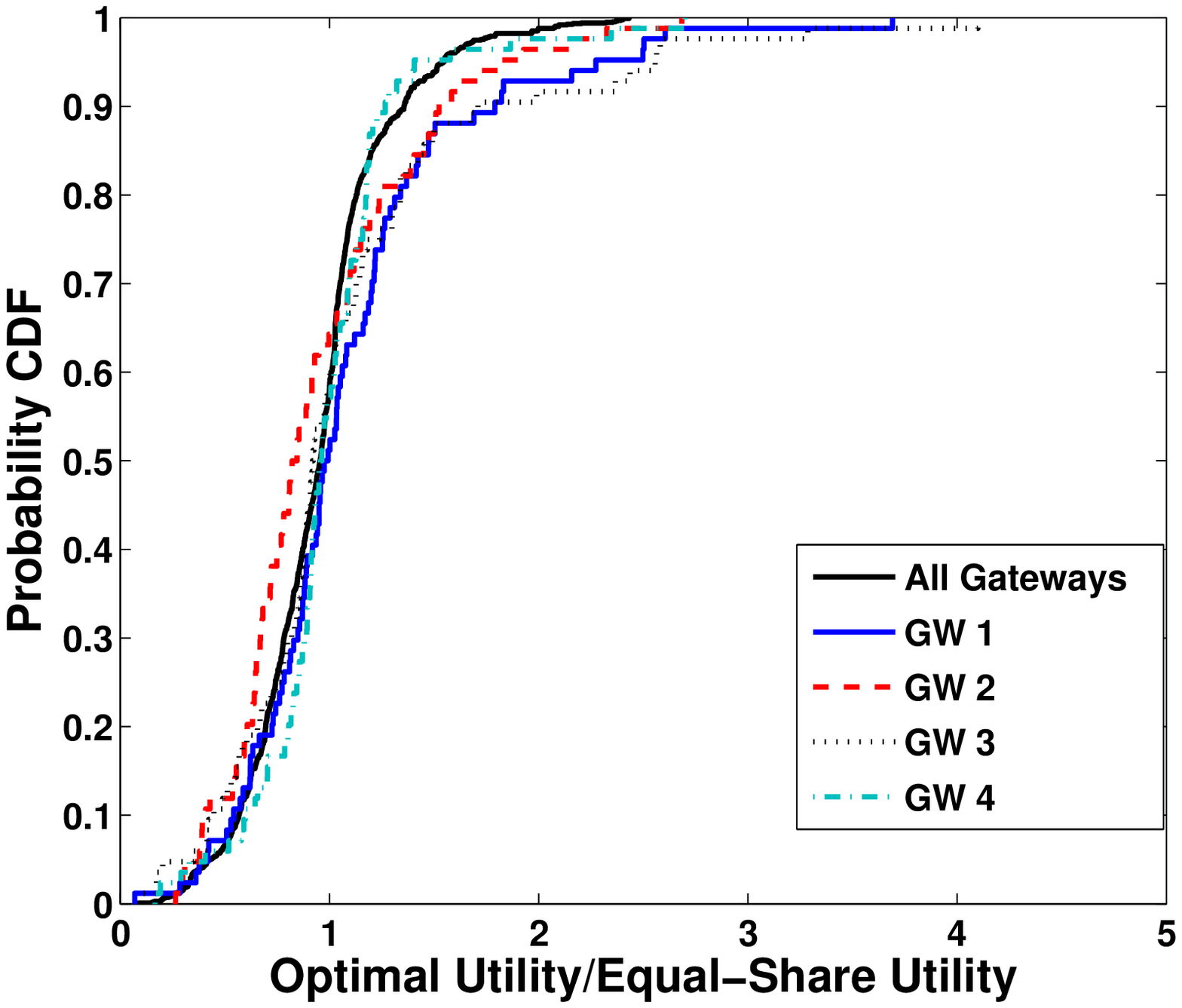}\label{fig:utility_compare}}
\vspace{-0.15in}
\caption{Rate variation, optimal credit budgets, and utilities for the gateways over one week.}
\vspace{-0.15in}
\end{figure*}

We suppose that credit-based sharing is enforced in the congested hours between 6pm and midnight, with half-hour timeslots.
Users at each gateway are assumed to make their credit spending decisions based on their probability of using four types of applications: streaming, social networking, file downloads, and web browsing. We use the utility functions
\begin{align*}
u_1(x) &= \frac{2(25x)^{1 - \alpha_1}}{1 - \alpha_1} \\
u_2(x) &= \frac{(25x)^{1 - \alpha_2}}{1 - \alpha_2} \\
u_3(x) &= \left(\frac{1}{\alpha_3 - 1} + \frac{(25x + 1)^{1 - \alpha_3}}{1 - \alpha_3}\right) \\
u_4(x) &= 15\left(\frac{1}{\alpha_4 - 1} + \frac{(25x + 1)^{1 - \alpha_4}}{1 - \alpha_4}\right)
\end{align*}
in (\ref{eq:utility}) to respectively model the utility received from each application, where $\left(\alpha_1, \alpha_2, \alpha_3, \alpha_4\right) = \left(0.7, 0.5, 0.2, 3\right)$. The probabilities $p_{it}^k$ of using each application are adapted from a recent measurement study of per-app usage over time for iOS, Android, Windows, and Mac smartphones and computers \cite{chung2011measurement}. Table \ref{tab:devices} shows the devices at each gateway.
\begin{table}
\label{tab:devices}
\caption{Devices at each gateway.}
\vspace{-0.05in}
\begin{tabular}{|c|c|c|c|c|}
\hline
Gateway & iPhones & Androids & Windows laptops & Mac laptops \\ \hline
1,4,9,13 & 1 & 1 & 1 & 1 \\
2,6,10,14 & 2 & 0 & 2 & 1 \\
3,7,11,15 & 1 & 1 & 1 & 2 \\
4,8,12,16 & 2 & 0 & 1 & 1 \\ \hline
\end{tabular}
\end{table}
We choose coefficients $\gamma_i(t)$ to be larger in the evening, as is consistent with the usage measurements in \cite{chung2011measurement}, and add random fluctuations to model heterogeneity between users and day-to-day variations in each gateway's behavior.

We assume a budget of $B = 160$ total credits, which each credit representing 1Mbps. The budget $b_{it}$ for each gateway $i$ is capped at 32 credits at any given time. In addition to the purchased bandwidth, we suppose that gateways send a random amount of traffic over the second tier, which is capped at the network capacity.
We simulate one week of credit redistributions and bandwidth allocations.

\subsection{Bandwidth Allocations}

{\bf Globally Optimal Solution:} We first compute the globally optimal rates, i.e., those that maximize (\ref{eq:isp_opt}), and show that the overall rate allocation is fair. To see this, we compute Jain's Index over the gateways' rates at each time, including second-tier traffic in Figure \ref{fig:rates_opt}. The Jain's Index is relatively low at some times, indicating that there is a large variation in gateways' rates. Thus, some gateways use little bandwidth in order to save credits and others spend a lot of credits and receive large bandwidth rates. Yet if we compute Jain's Index for all gateways' \emph{cumulative} usage over time, we see in Figure \ref{fig:rates_opt} that the index quickly converges to 1. The gateways receive comparable cumulative rates, as is consistent with the fairness property of Prop. \ref{prop:bound}.

The large variability in gateway allocations at a given time can be seen more clearly in Figure \ref{fig:budgets_opt}, which shows the budgets of four representative gateways over time. All four save credits at some times in order to spend at other times. This time flexibility significantly improves the overall utility over equal sharing: the gateways' total achieved utility increases by 29.7\% relative to the equal sharing allocation of each gateway receiving 1.25Mbps at all times.

Figure \ref{fig:utility_compare} shows the cumulative density function (CDF) of the ratio of gateway utilities under credit allocation and equal sharing at different times. We plot the CDF for all gateways and times, as well as individual CDFs for the gateways shown in Figure \ref{fig:budgets_opt}. All of the ratio distributions are comparable, indicating that credit-based allocation benefits all gateways' utilities. While the gateways reduce their utility nearly half of the time, the utility more than doubles in some periods.

{\bf Online Solution:} We next compare the globally optimal utilities with those obtained when the gateways follow Algorithm \ref{alg:allocate}. To perform the credit estimation, we use four scenarios, in which all other gateways are assumed to use only streaming, only social networking, etc. Each gateway assumes (falsely) that the other gateways' $\gamma_i(t)$ coefficients are the same, and the probabilities $p_\sigma$ of each scenario are initialized to be uniform. After learning the scenario distribution for the simulation's first four days, the algorithms recovers 84.7\% of the optimal utility for the remaining three days. Since this result is achieved with only four scenarios, Algorithm \ref{alg:allocate} is practically effective in achieving near-optimal rates.

\begin{figure*}
\centering
\vspace{-0.1in}
\subfigure[Jain's Index for gateway rates.]{\includegraphics[width = 0.4\textwidth]{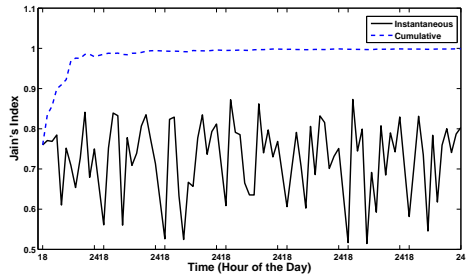}\label{fig:rates_dist}}
\subfigure[Credit budgets for representative gateways.]{\includegraphics[width = 0.4\textwidth]{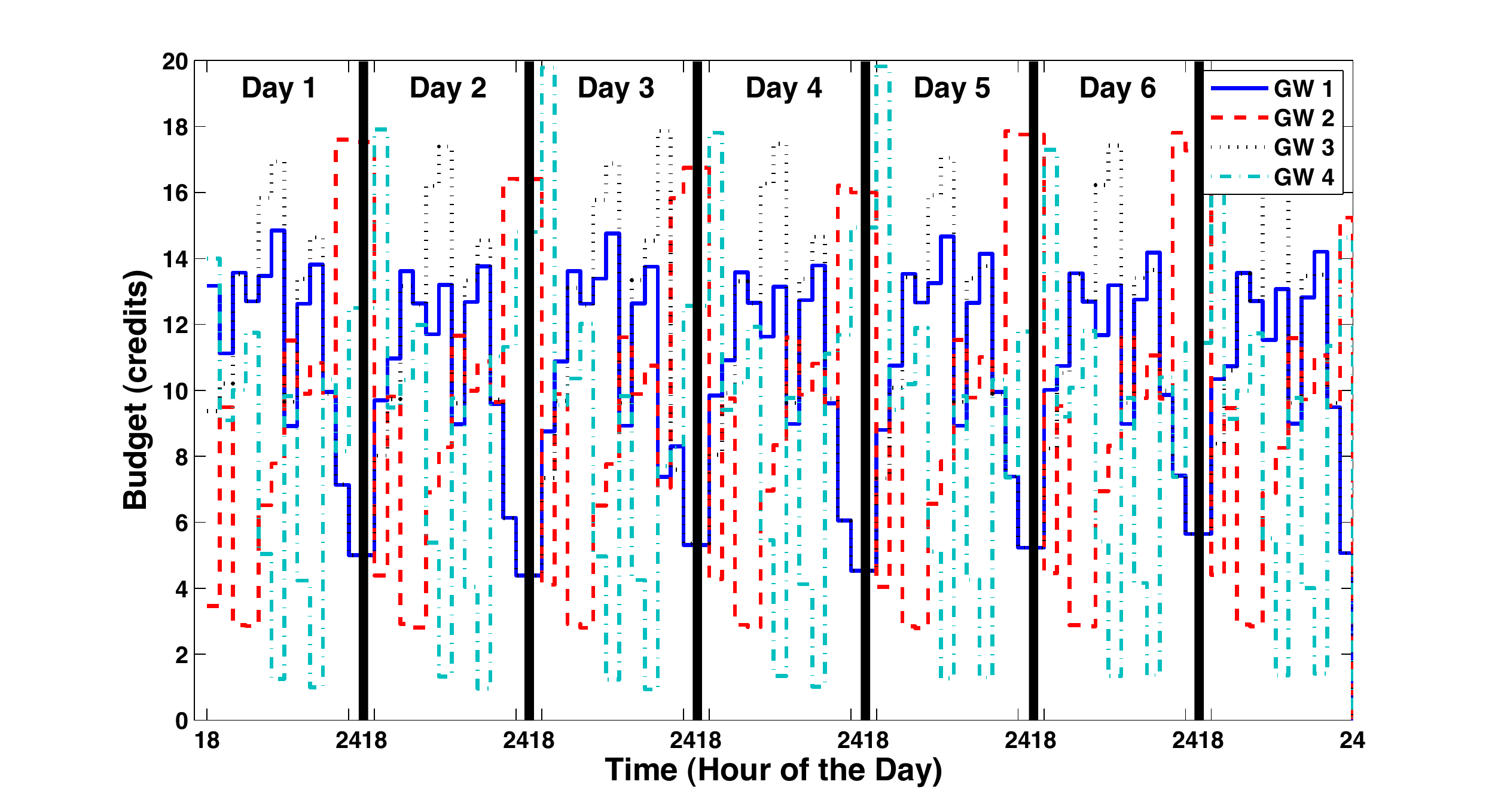}\label{fig:budgets}}
\vspace{-0.1in}
\caption{Rate variation and optimal credit budgets with the online algorithm for the gateways over one week.}
\vspace{-0.15in}
\label{fig:dist}
\end{figure*}

As with the optimal solution, at any given time gateways' rates can be very different: Jain's indices in Figure \ref{fig:rates_dist} for all gateways' usage at instantaneous times can be quite low. However, all gateways achieve similar cumulative rates: Jain's index of the cumulative rates quickly converges to 1. Indeed, Figure \ref{fig:budgets} shows the budgets of four representative gateways, indicating that, as with the optimal solution (Figure \ref{fig:budgets_opt}), the gateways vary their spending with time. Thus, incentivizing gateways to delay some of their usage significantly improves users' overall satisfaction and utility.

\section{Related Work}\label{sec:related}

Using pricing to manage network congestion is a long-studied research area \cite{Sen}.
Our work differs in that we target broadband users on flat-fee service plans,
prompting us to use virtual pricing instead of having users pay extra fees for prioritized access.
Related to our work is \cite{lee2011token}, which proposes a scheme for users to access higher-quality service
by spending tokens, but their work is mostly theoretical. 
We present a complete solution, from algorithms to implementation, for a specific problem of peak-hour broadband access.

From a systems perspective, there has been much recent work on developing smart home gateways with plain Linux/Windows or open-source router software
such as OpenWrt.
Smart home gateways have been used for network measurement \cite{sundaresan2011,patro2013}, providing intuitive
interfaces for home network management \cite{mortier2012homework,yang2010eden} and better QoS provisioning \cite{palazzi2010openwrt,gkantsidis2009}.
However, we are not aware of any work in coordinating bandwidth usage across households.
We also develop our own incoming rate limiting tool, as off-the-shelf tools (e.g., Linux $\mathtt{tc}$) are insufficient for our application.
The Congestion Manager (CM) project \cite{balakrishnan1999} shares similar goals of reducing congestion 
at the network edge, but we approach the problem by incentivizing users to reduce unneeded usage with a virtual currency scheme,
while CM provides an API for client and server applications to adapt to varying network conditions. CM thus requires sender-side support.

Receiver-side rate control is mostly done through
explicitly controlling the receive window \cite{kalampoukas1998} or the receive socket buffer \cite{semke1998}.
They have been applied in implementing low-priority transfers
\cite{key2004} and prioritizing traffic \cite{spring2000,im2013}.
Compared to these approaches, our solution does not require modifying client devices or tracking both RTT and the number
of active connections. It also avoids interfering with Linux's own buffer autotuning mechanism \cite{fisk2001}, which is
crucial for connections with high bandwidth-delay products.
Our approach of implicit receive window control is most similar to that of Trickle \cite{eriksen2005},
but the goals are different.
Trickle is designed for users without administrative privileges to \emph{voluntarily} rate
limit their applications, while we are interested in imposing mandatory rate limits transparent to users.
For rate limiting across multiple devices, \cite{eriksen2005} proposes a distributed architecture where
a centralized scheduler (e.g., the gateway) coordinates rate limiting over a home network.
Our solution, in using a split-connection proxy, allows rate limiting to be done solely inside the home gateway
to avoid the overhead of distributed coordination.

\section{Concluding Remarks}\label{sec:conclusion}

In this paper, we propose to solve peak-hour broadband network congestion problems by pushing congestion management
to the network edge.
Our solution performs a two-level bandwidth allocation: in Level 1, gateways purchase bandwidth on a shared link using virtual credits, and in Level 2 they divide the purchased bandwidth among their apps and devices. We show analytically that our credit distribution scheme yields a fair bandwidth allocation across gateways and describe our implementation of the bandwidth purchasing and app prioritization on commodity wireless routers. Our implementation can successfully enforce app priorities and increase users' satisfaction. Finally, we simulate the behavior of sixteen gateways sharing a single link. We show that our algorithm yields a fair bandwidth allocation that significantly improves user utility relative to a baseline equal-sharing scheme.

By implementing congestion management at the network edge, we obtain a decentralized, personalized solution that respects user privacy and requires minimal support from ISP infrastructure and user devices. Since users make their own decisions regarding credit spending and bandwidth allocation, our incentive mechanisms \emph{empower users} to moderate their demand so as to limit network congestion. While we implement this solution for cable networks, our methodology is applicable to other access technologies, \eg cellular,
that involve shared medium access. Such technologies, wireless and wired, will increasingly need new congestion management mechanisms as user demand for bandwidth continues to grow.

\bibliographystyle{IEEEtran}
\small
\bibliography{myref}

\newpage

\normalsize
\appendices

\section{Proof of Lemma \ref{lem:conserve}}\label{sec:proof1}

We proceed by induction: at time $t = 0$, clearly the sum of gateways' budgets $\sum_i b_{i0} = B$ from the budget initialization. Supposing that $\sum_{i = 1}^n b_{it} = B$ at time $t$, we then calculate
\begin{align*}
\sum_{i = 1}^n b_{i,t + 1} &= \sum_{i = 1}^n \left(b_{it} - x_{it} + \sum_{j \neq i}\frac{x_{jt}}{n - 1}\right) \\
&= B - \sum_{i = 1}^n x_{it} + \sum_{i = 1}^n \frac{(n - 1)x_{it}}{n - 1} = B.
\end{align*}

\section{Proof of Lemma \ref{lem:equal}}

We first note that (\ref{eq:credits}) is equivalent to the statement that
\begin{equation*}
b_{it} = b_{is} + \sum_{\tau = s}^{t - 1}\left(-x_{i\tau} + \sum_{j\neq i} x_{j\tau}/(n - 1)\right).
\end{equation*}
Then if $b_{is} = b_{it}$ for all gateways $i$, we obtain the system of equations
\begin{equation}
\sum_{\tau = s}^{t - 1} x_{i\tau} = \sum_{\tau = s}^{t - 1} \sum_{j\neq i}\frac{x_{j\tau}}{n - 1}.
\label{eq:system}
\end{equation}
It suffices to show that (\ref{eq:system}) implies the proposition.

We proceed by induction on $n$. If $n = 2$, then clearly (\ref{eq:system}) is exactly our desired result, since $n - 1 = 1$. We now suppose that the proposition holds for $n = m$ and show that it holds for $n = m + 1$. From (\ref{eq:system}), we have
\begin{equation*}
\sum_{\tau = s}^{t - 1} x_{1\tau} = \sum_{\tau = s}^{t - 1} \sum_{j = 2}^n\frac{x_{j\tau}}{n - 1}.
\end{equation*}
Substituting this equality into (\ref{eq:system}) for $i > 1$, we have for all such $i$,
\begin{equation*}
\sum_{\tau = s}^{t - 1} x_{i\tau} = \sum_{\tau = s}^{t - 1} \sum_{j = 2}^n\frac{x_{j\tau}}{(n - 1)^2} + \sum_{\tau = s}^{t - 1} \sum_{j\neq i, j > 1}\frac{x_{j\tau}}{n - 1}.
\end{equation*}
Thus, we have upon rearranging that
\begin{equation*}
\left(1 - \frac{1}{(n - 1)^2}\right)\sum_{\tau = s}^{t - 1} x_{i\tau} = \left(\frac{1}{(n - 1)^2} + \frac{1}{n - 1}\right)\sum_{\tau = s}^{t - 1} \sum_{j\neq i, j > 1}x_{j\tau}.
\end{equation*}
Simplifying, we obtain
\begin{equation*}
\sum_{\tau = s}^{t - 1} x_{i\tau} = \sum_{\tau = s}^{t - 1} \sum_{j\neq i, j > 1}\frac{x_{j\tau}}{n - 2}
\end{equation*}
for all $i > 1$. By induction, this implies that $\sum_{\tau = s}^{t - 1} x_{j\tau} = \sum_{\tau = s}^{t - 1} x_{k\tau}$ for all $j,k > 1$, and the proposition follows upon solving for $\sum_{\tau = s}^{t - 1} x_{1\tau}$.

\section{Proof of Proposition \ref{prop:bound}}

We first show that given a distribution of budgets $\left\{b_{it}\right\}$ at a fixed time $t$, there exists a set of gateway spending decisions $\left\{x_{it}\right\}$ such that $b_{i,t + 1} = B/n$ for all gateways $i$. Suppose that each gateway $i$ spends $x_{it} = b_{it}(n - 1)/n$ credits at time $t$. Then Lemma \ref{lem:conserve}'s budget conservation allows us to conclude that gateway $i$'s budget at time $t + 1$ is
\begin{equation*}
b_{i,t + 1} = b_{it} - \frac{b_{it}(n - 1)}{n} + \sum_{j\neq i}\frac{b_{jt}(n - 1)}{n(n - 1)} = \sum_{i = 1}^n \frac{b_{it}}{n} = \frac{B}{n}.
\end{equation*}
We now observe that since each $b_{i0} = B/n$, we can apply Lemma \ref{lem:equal} to conclude that
\begin{equation*}
\sum_{s = 0}^{t + 1} x_{is} = \sum_{s = 0}^{t} x_{is} + \frac{b_{it}(n - 1)}{n} = \sum_{s = 0}^{t} x_{js} + \frac{b_{jt}(n - 1)}{n} = \sum_{s = 0}^{t + 1} x_{js}
\end{equation*}
for all gateways $i$ and $j$. We then rearrange this equation to find the first part of the proposition:
\begin{equation*}
\left|\sum_{s = 0}^{t} x_{is} - \sum_{s = 0}^{t} x_{js}\right| = \left|b_{jt} - b_{it}\right|\frac{n - 1}{n} \leq \frac{B(n - 1)}{n}.
\end{equation*}
The time average follows immediately upon dividing by $t$ and taking limits as s$t\rightarrow\infty$.

\section{Proof of Proposition \ref{prop:limit}}

To prove the first part of the proposition, we note that if each $x_{it} = 0$, then (\ref{eq:credits}) yields
\begin{equation*}
b_{i,t + 1} = b_{it} - x_{it} + \sum_{j \neq i}\frac{x_{jt}}{n - 1} \leq \frac{B}{n-1} + b_{it}\frac{n-2}{n - 1} - x_{it},
\end{equation*}
where the inequality comes from each gateway's budget constraint $\sum_{j \neq i} x_{jt} \leq \sum_{j \neq i} b_{jt} = B - b_{it}$. Thus, at time $t + 1$, we have
\begin{align*}
b_{i,t + 1} &= \sum_{\tau = 0}^t \left(\frac{B}{n - 1} - x_{i\tau}\right)\left(\frac{n - 2}{n - 1}\right)^\tau + \frac{B}{n}\left(\frac{n - 2}{n - 1}\right)^{t + 1} \\
&\leq \frac{B}{n}\alpha^{t + 1} + B\left(1 - \alpha^{t + 1}\right) - \epsilon\sum_{\tau = 1}^{\left\lfloor\frac{t + 1}{p}\right\rfloor} \alpha^{p\tau} \\
&= \frac{B}{n}\alpha^{t + 1} + B\left(1 - \alpha^{t + 1}\right) - \epsilon\left(\frac{\alpha^p - \alpha^{p\left(1 + \left\lfloor\frac{t + 1}{p}\right\rfloor\right)}}{1 - \alpha^p}\right)
\end{align*}
as desired, using the fact that $\sum_{\tau = s}^{s + n} x_{i\tau} \geq \epsilon$ at any time $s$. We obtain (\ref{eq:limit}) by taking $t\rightarrow\infty$, substituting for $\alpha = \frac{n - 2}{n - 1}$, and simplifying.

To prove the second part of the proposition, suppose that gateways $i$ and $k$ both have zero budgets at time $t + 1$, i.e., $b_{i,t + 1} = b_{k,t + 1} = 0$, but that $b_{it} > 0$. Since each $b_{i0} = B/n > 0$, such a time $t $ must exist. But then from (\ref{eq:credits}), $b_{i, t + 1} = b_{it} - x_{it} + \sum_{j\neq i} x_{jt}/(n - 1) = 0$, and since each $x_{jt} \geq 0$, we have $x_{it} = b_{it} > 0$. But then $b_{k,t + 1} = b_{kt} - x_{kt} + \sum_{j\neq k} x_{jt}/(n - 1)$, and since $x_{kt} \leq b_{kt}$, we have $b_{k, t + 1} > 0$, which is a contradiction. Thus, at most one gateway can have zero budget in any given time period.

\section{Proof of Proposition \ref{prop:limit_num}}

We first note that at each time $t < s$,
\begin{align*}
\sum_{i = 1}^m b_{i,t + 1} &\leq \sum_{i = 1}^m b_{it} + \sum_{j > i}\frac{x_{it}}{n - 1} \\
&\leq \sum_{i = 1}^m b_{it} + \frac{B - \sum_{i = 1}^m b_{it}}{n - 1} \\
&= \frac{B}{n - 1} + \left(\frac{n - 2}{n - 1}\right)\sum_{i = 1}^m b_{it}.
\end{align*}
An inductive argument then shows that
\begin{equation*}
\sum_{i = 1}^m b_{i,s} \leq \frac{B}{n - 1}\left(\sum_{\tau = 0}^{s - 1} \left(\frac{n - 2}{n - 1}\right)^s\right) + \left(\frac{n - 2}{n - 1}\right)^s \sum_{i = 1}^m b_{i0}.
\end{equation*}
Expanding the sums and subtracting $\sum_{i = 1}^m b_{i0}$ then yields the proposition.

\section{Proof of Proposition \ref{prop:nash}}\label{sec:prooflast}

Suppose that $\left\{x_{it}^\ast\right\}$ solve (\ref{eq:isp_opt}), and let $\lambda_{it}$ denote the corresponding Lagrange multiplier for the constraint $0\leq x_{it} \leq b_{it}$, with $\nu_{it}$ the multiplier for the constraint $x_{it} \geq 0$. Since the $U_{it}$ are strictly concave, it suffices to show that these multipliers satisfy the Karush-Kuhn-Tucker conditions for (\ref{eq:gateway_opt}), augmented by all gateways' constraints:
\begin{equation*}
\max_{x_{it}} \sum_{t = s}^{s + T} U_{it}\left(x_{it}\right),\;\;{\rm s.t.}\;0\leq x_{it} \leq b_{it},\;\forall i,t.
\end{equation*}
Since the budget constraints $0\leq x_{it} \leq b_{it}$ are identical to those of (\ref{eq:isp_opt}), it suffices to show that 
\begin{equation}
\frac{d U_{it}}{dx_{it}} - \sum_{j = 1}^n\sum_{\tau = t}^{s + T} \lambda_{i\tau} + \sum_{j\neq i}\sum_{\tau = t}^{s + T - 1} \frac{\lambda_{j\tau}}{n - 1} + \nu_{it} = 0,
\label{eq:KKT}
\end{equation}
\newpage
where we use (\ref{eq:budget_new}) to sum over the appropriate multipliers $\lambda_{i\tau}$. However, this equation is just one of the KKT conditions for (\ref{eq:isp_opt}): the only change between (\ref{eq:isp_opt}) and (\ref{eq:gateway_opt}) is the addition of utility terms $U_{jt}(x_{jt})$, which are additively decoupled from gateway $i$'s spending decisions $x_{it}$. Thus, (\ref{eq:KKT}) must be satisfied by the $x_{it}^\ast$ and multipliers $\lambda_{it}$, $\nu_{it}$. Each gateway $i$ is thus optimizing its own utility, given other gateways' credit spending decisions $x_{jt}^\ast$.

\section{Rate Control by Controlling Buffer Reads}\label{sec:buffer_model}
\begin{figure}
\centering
\includegraphics[width=0.35\textwidth]{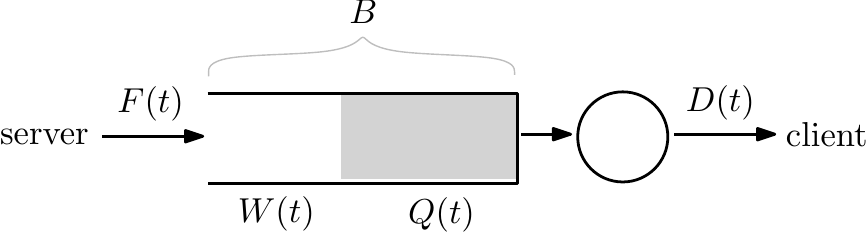}
\caption{Receive buffer model.}
\label{fig:onebuf}
\end{figure}

Consider the model in Figure \ref{fig:onebuf}:
the queue is the proxy's receive buffer,
$B$ is the receive buffer size (it can change with time due to Linux's buffer autotuning
mechanism but it does not concern us here),
and at time $t$, $F(t)$ is the fill rate (sending rate, which the proxy cannot directly control),
$D(t)$ is the drain rate (how frequent the proxy issues $\mathtt{recv()}$'s), 
$Q(t)$ is the queue length, and $W(t)=B-Q(t)$ is the advertised window size.

Suppose updates happen at intervals of $\Delta t$, then the window update equation is
\begin{align}
W(t+\Delta t) &= W(t) + \bigl[ D(t) - F(t) \bigr]^+ \Delta t
\end{align}
and taking a fluid approximation by setting $\Delta t \to 0$, we have
\begin{align}
\dot{W}(t) &= \bigl[ D(t) - F(t) \bigr]^+\label{fluid}.
\end{align}

Our goal of rate limiting is equivalent to getting $F(t)=R$ for large enough $t$
through controlling $D(t)$
By setting $D(t)=R$ at all $t$, 
it is not difficult to verify from Eq.\ \eqref{fluid} that at equilibrium\footnote{Note that if we throttle
a connection through TCP flow control, a static equilibrium can indeed be achieved because the rate is now
limited by the receive window, rather than self-induced congestion, i.e., the usual sawtooth $W(t)$
time evolution does not happen anymore.}
we have $F^*(t) = R$ and $W^*(t) = R\times \mathtt{RTT}$.

\end{document}